\documentclass[useAMS,usenatbib,usegraphicx,letterpaper]{mn2e}

\usepackage{multirow} 
\usepackage{epsfig}
\usepackage{subfigure}

\newcommand{\etal}{et al.}

\def\mnras{MNRAS}
\def\apj{ApJ}
\def\apjl{ApJ Letters}
\def\apjs{ApJS}
\def\aap{A\&A}
\def\aj{AJ}
\def\nat{Nature}
\def\aapr{A\&A Review}

\title[An outflowing X--ray corona in RBS~1124?] {Does the X--ray
  emission of the luminous quasar RBS~1124 originate in a mildly
  relativistic outflowing corona?}  \author[G.\ Miniutti \etal]
      {G. Miniutti$^{1}$\thanks{gminiutti@laeff.inta.es},
        E. Piconcelli$^2$, S. Bianchi$^3$, C. Vignali$^{4,5}$ and E. Bozzo$^6$
        \\ \\ $^1$ Centro de Astrobiologia (CSIC--INTA); LAEFF, P.O:
        Box 78, E-28691, Villanueva de la Ca\~nada, Madrid, Spain \\
        $^2$ INAF--Osservatorio Astronomico di Roma, Via Frascati 33,
        I--00040, Monteporzio Catone, Italy \\ $^3$ Dipartimento di
        Fisica, Universit\`a degli Studi Roma Tre, via della Vasca
        Navale 84, 00146 Roma, Italy \\ $^4$ Dipartimento di
        Astronomia, Universit\`a degli Studi di Bologna, Via Ranzani
        1, 40127 Bologna, Italy \\ $^5$ INAF-–Osservatorio Astronomico
        di Bologna, Via Ranzani, 1, 40127 Bologna, Italy \\ $^6$
        INTEGRAL Science Data Center (ISDC), Science Data center for
        Astrophysics, Ch. d'Ecogia 16, CH--1290 Versoix (Ge),
        Switzerland }

\pagerange{\pageref{firstpage}--\pageref{lastpage}}
\pubyear{2001}

\usepackage{times}

\begin{document}

\label{firstpage}

 \maketitle

\begin{abstract}
We have observed the luminous ($L_{2-10~keV} \simeq 6\times
10^{44}$~erg~s$^{-1}$) radio--quiet quasar RBS~1124 ($z=0.208$) with
{\it Suzaku}. We report the detection of a moderately broad iron (Fe)
line and of a weak soft X--ray excess. The X--ray data are very well
described by a simple model comprising a power law X--ray continuum
plus its reflection off the accretion disc. If the inner disc radius
we measure ($r_{\rm in}\leq 3.8$ gravitational radii) is identified
with the innermost stable circular orbit of the black hole spacetime,
we infer that the black hole powering RBS~1124 is rotating rapidly
with spin $a\geq 0.6$. The soft excess contribution in the 0.5--2~keV
band is $\sim$15 per cent, about half than that typically observed in
unobscured Seyfert 1 galaxies and quasars, in line with the low disc
reflection fraction we measure ($R_{\rm disc}\simeq 0.4$). The low
reflection fraction cannot be driven by disc truncation which is at
odds not only with the small inner disc radius we infer but, most
importantly, with the radiatively efficient nature of the source
($L_{\rm Bol} / L_{\rm Edd} \simeq 1$). A plausible explanation is
that the X--ray corona is the base of a failed jet (RBS~1124 being
radio--quiet) and actually outflowing at mildly relativistic
speeds. Aberration reduces the irradiation of the disc, thus forcing a
lower than standard reflection fraction, and halves the inferred
source intrinsic luminosity, reducing the derived Eddington ratio from
$\simeq 1$ to $\simeq 0.5$. A partial covering model provides a
statistically equivalent description of the 0.3--10~keV data, but
provides a worse fit above 10~keV. More importantly, its properties
are not consistent with being associated to the Fe emission line,
worsening the degree of self--consistency of the model. Moreover, the
partial covering model implies that RBS~1124 is radiating well above
its Eddington luminosity, which seems unlikely and very far off from
previous estimates.
\end{abstract}


\section{Introduction}

The X-ray spectrum of unobscured (type I) Active Galactic Nuclei (AGN)
is generally dominated by a power law component which most likely
comes from Compton up--scattering of the soft UV/EUV photons from the
accretion disc in hot active regions (the so--called X--ray
corona). This simple spectral description is however almost
ubiquitously modified by a soft X--ray excess below 1--2~keV
(e.g. Pounds et al. 1987), by the presence of a narrow iron (Fe)
emission line at $\sim$6.4~keV (e.g. Bianchi et al. 2007), and by
partially ionized absorbers in the line of sight (the so--called warm
absorbers, see e.g. Blustin et al. 2005). Broad Fe emission lines are
also sometimes observed and generally attributed to X--ray reflection
of the primary power law continuum off the accretion disc (e.g. Nandra
et al. 2007; Fabian \& Miniutti 2009). While the narrow Fe line origin
is not highly debated and is most likely associated with reflection
from distant matter (such as the obscuring material of AGN unification
schemes) and the warm absorber properties can be inferred from the
analysis of high resolution spectra, the nature of the soft excess
emission is not yet well established.

The soft excess typically rises smoothly above the extrapolation of
the 2--10~keV continuum and its shape is suggestive of low temperature
($kT\sim 150$~eV) Comptonization of the soft disc photons, a much
lower electron temperature than that generally invoked to account for
the high energy power law component ($kT\sim 100$~keV). However the
``soft excess temperature'' is remarkably uniform in radiatively
efficient sources despite a large range in black hole mass and hence
disc temperatures, casting doubts on this interpretation (see
e.g. Walter \& Fink 1993; Czerny et al. 2003; Gierlinski \& done 2004;
Middleton et al. 2007; Miniutti et al. 2009a). The uniform temperature
of the soft excess may have a natural explanation if the soft excess
is due to atomic rather than thermal processes. Partially ionized
material may produce a uniform soft excess in a relatively wide
ionization state range because of excess opacity in the 0.7--2~keV
range with respect to neighbouring energy bands. This simple idea
leads to two main competing atomic--based interpretations of the soft
excess in AGN, namely absorption by partially ionized gas or X--ray
reflection from the accretion disc material (see e.g. Gierlinski \&
Done 2004; Crummy et al. 2006).

High resolution spectroscopy of the soft X--ray spectra of AGN does
indeed reveal a series of absorption features, but the associated gas
is not able to account for the soft excess (e.g. Blustin et al. 2005
and references therein). Hence, for the absorption interpretation to
be consistent with the available data, the predicted sharp absorption
features have to be smoothed out by very large velocity smearing such
as those expected in fast outflows and winds. However, when physical
models are considered, absorbing winds/outflows are able to produce
significant soft excesses only by also predicting relatively sharp
absorption features in the soft X--rays which are not consistent with
the data, casting doubts on the overall interpretation or,
alternatively, calling for more complex physical models (Sim et
al. 2008; Schurch, Done \& Proga 2009). On the other hand, the sharp
features associated with the X--ray reflection spectrum are naturally
smoothed out by the large orbital velocities of the inner accretion
disc, so that no additional physics besides the standard disc/corona
structure is required to explain the data in the framework of the disc
reflection interpretation of the soft excess (Fabian \& Miniutti
2009). The recent detections of broad Fe L and K emission lines and of
a reflection time lag in the NLS1 galaxy 1H~0707--495 favour the
reflection interpretation in this source, possibly indicating that
this is indeed the right interpretation for the soft excess in most
cases (Fabian et al. 2009; see also Fabian et al. 2004).

RBS~1124 ($z=0.208$, a.k.a. RX~J1231.6+7044) is a broad line quasar
(QSO) with H$\beta$~FWHM of $4.26\pm 1.25\times 10^3$~km~s$^{-1}$ and
optical luminosity $\lambda L_{5100} \simeq 1.26\times
10^{44}$~erg~s$^{-1}$ (Grupe et al. 2004). The 5100\AA\ luminosity,
combined with the H$\beta$ width, can be used to infer a black hole
mass of $\simeq 1.8\times 10^8~M_\odot$ (e.g. by using the
relationships in Vestergaard \& Peterson 2006). The bolometric
luminosity has been estimated by Grupe et al. (2004) from a combined
fit to the optical--UV and X--ray data and is $L_{\rm Bol} \simeq 3.4
\times 10^{45}$~erg~s$^{-1}$ which, together with the black hole mass
estimate, implies that RBS~1124 is radiatively efficient with an
Eddington ratio of $\simeq 0.15$. Note however that, as pointed out
in Grupe et al. (2004), the derived bolometric luminosity has to be
considered as an estimate only given the lack of EUV data.

In the X--rays, RBS~1124 was detected by ROSAT, and is in the ROSAT
All--sky Bright Source Catalogue (Voges et al. 1999) with a RASS
0.2--2~keV flux of $\sim 7.6\times 10^{-12}$~erg~cm$^{-2}$~s$^{-1}$
and a spectral shape consistent with a simple power law model absorbed
by the Galactic column density (Grupe, Thomas \& Beuermann 2001). It
is also present in the first release of the {\it XMM--Newton} Slew
Survey Catalogue XMMSL1 (Saxton et al. 2008) with a 0.2--2~keV
EPIC--pn count rate of $\sim 3.9$~cts~s$^{-1}$, corresponding to a
flux of $\sim 6.5\times 10^{-12}$~erg~cm$^{-2}$~s$^{-1}$ in the same
band (assuming an absorbed power law model with N$_{\rm H}=1.5\times
10^{20}$~cm$^{-2}$ and $\Gamma = 2.38$, as obtained from the RASS by
Grupe et al. 2001).

The relatively high Eddington ratio of RBS~1124 suggests that its
X--ray spectrum may share a series of common properties with the class
of Seyfert 1 galaxies, and in particular the source may represent a
rare opportunity to observe a soft excess at very high X--ray
luminosities. We then decided to observe the source in the widest
possible X--ray band by using the Japanese (JAXA/NASA) X--ray
observatory {\it Suzaku} which is sensitive enough (for the given
source brightness) to provide accurate spectral information in the
$\sim$0.3--30~keV band. Here we present results from a $\sim$86~ks
{\it Suzaku} observation of RBS~1124, and we also present for the first
time results from {\it Swift}/XRT observations of the source.

\section{Observations and data analysis}

We have observed RBS~1124 with {\it Suzaku} on 2007 April 14th
(observation ID: 702114010) obtaining a net exposure of 86~ks
($\sim$~140~ks total exposure). The source was placed at the HXD
nominal position. Products for the three available XIS units (the two
front--illuminated CCD XIS~0 and XIS~3 and the back--illuminated CCD
XIS~1, hereafter FI and BI CCD) and for the HXD PIN have been
extracted following the appropriate version of the Suzaku Data
Reduction Guide\footnote{Available at
  http://heasarc.gsfc.nasa.gov/docs/suzaku/analysis/abc/}. Our
analysis is based on data processing version 2.0.6.13 and is carried
out with the HEASOFT 6.6.2 software and the {\it Suzaku} {\sc{caldb}}
released on 2009--04--03. The XIS data have been analysed starting
from the unscreened event files and by applying the {\sc{xispi}} tool
and standard screening criteria to produce screened cleaned event
files (see Chapter 6 of the Suzaku Data Reduction Guide). Appropriate
{\sc{rmf}} and {\sc{arf}} files have been generated with the
{\sc{xisrmfgen}} and {\sc{xissimarfgen}} tools. As for the HXD/PIN,
besides standard reduction, we have accounted for the Cosmic X--ray
background (CXB) by simulating the CXB contribution to the PIN
background according to the recipes in Chapter~7 of the Suzaku Data
Reduction Guide. Given that the CXB model in the hard energy band
still suffers from a $\sim$10 per cent uncertainty (Roberto Gilli,
private communication), we included a 10 per cent systematic
uncertainty to the CXB spectrum and we then added the corrected CXB
spectrum to the instrumental non X--ray background (NXB) provided by
the {\it Suzaku} team to which a 1.4 per cent systematic uncertainty
was also added (following Fukazawa et al. 2009). We use the
appropriate PIN response matrix for the epoch of the observation at
the HXD nominal position.

RBS~1124 was also observed by {\it Swift}/XRT on four occasions from
2007--05--24 to 2007--07--21, i.e. starting about 40 days after the
{\it Suzaku} observation (observation ID 00036542001, 00036542002,
00036542003, and 00036542004) with exposures in the range of
0.6--5.3~ks. {\it Swift}/BAT data were collected in survey mode and
cannot be analysed in this work due to the limited public availability
of the analysis software. In all observations, XRT data were accumulated in
Photon Counting (PC) mode and we present here the XRT/PC data for the
first time. We processed these data with the {\sc
  xrtpipeline} (v.0.12.1), and we applied standard filtering and
screening criteria.  Source and background light curves and spectra
were extracted by selecting standard event grades of 0--12, and using
circular extraction regions with radii of 20 pixels. We created
exposure maps by using the {\sc xrtexpomap} task, and used the latest
spectral redistribution matrices available from the {\it Swift} {\sc caldb}.
Ancillary response files, accounting for different extraction regions,
vignetting and PSF corrections, were generated by using the {\sc
  xrtmkarf} task. As discussed in the text, given that the single
spectra are all consistent with the same shape and X--ray flux, we
have co--added the four available XRT/PC exposures to produce a single
spectrum with an exposure of $\sim$14~ks.
 
Spectral analysis is performed with the {\sc xspec} v.12.5 software. All
{\it Suzaku} spectra are grouped to a minimum of 50 counts per bin (20
for the XRT/PC spectra) to ensure the applicability of $\chi^2$
statistics. We report 90 per cent errors ($\Delta\chi^2 = 2.706$ for 1
parameter) unless stated otherwise. Fluxes below 10~keV are those
observed with the {\it Suzaku} XIS~0 detector (or with the XRT/PC, when
stated). We adopt a standard $\Lambda$CDM cosmology with
H$_0=70$~km~s$^{-1}$, $\Omega_\Lambda = 0.73$, and $\Omega_{\rm M} =
0.27$.

\section{The 2--10~keV spectrum and Fe emission line}

\begin{figure}
\begin{center}
\includegraphics[width=0.35\textwidth,height=0.45\textwidth,angle=-90]{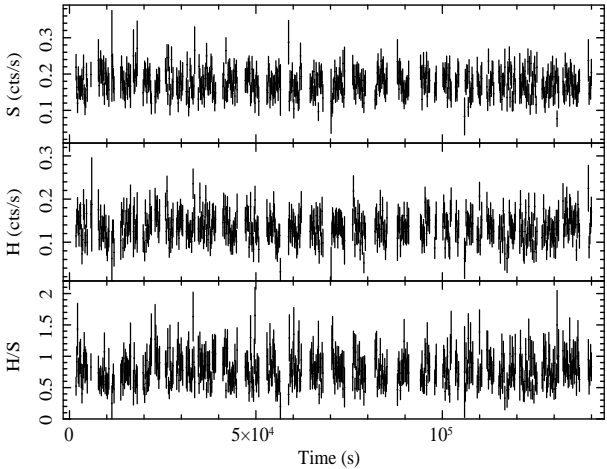}
\caption{The background subtracted XIS~3 light curves with 200~s time
  bins. In the two top panels we show the soft (S) 0.5--2~keV and hard
  (H) 2--10~keV light curves, while the bottom panel displays their ratio
  (simply defined as H/S).}
\label{lcs}
\end{center}
\end{figure}

The 0.5--10~keV XIS~0 light curve of RBS~1124 is shown in Fig.~\ref
{lcs} together with the hard to soft counts ratio light curve (H/S where H and S
are the count rates in the 2--10~keV and 0.5--2~keV band
respectively). No significant flux nor spectral variability is present
throughout the exposure. We then proceed with the analysis of the
time--averaged properties of the source.

\begin{figure}
\begin{center}
\includegraphics[width=0.35\textwidth,height=0.45\textwidth,angle=-90]{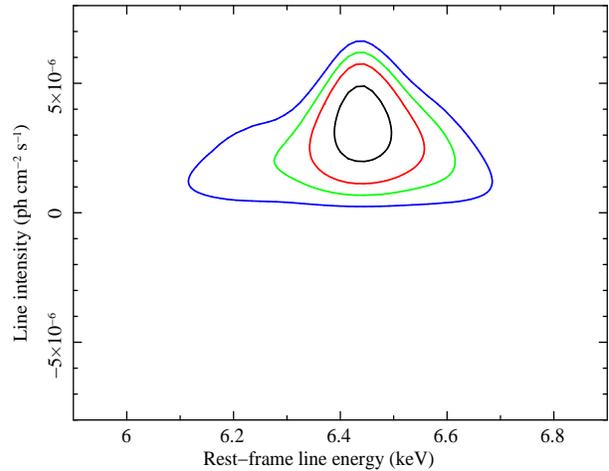}
\caption{$\Delta\chi^2$ contours when a Gaussian filter is run through
  the data modelled with a simple absorbed power law. No other
  features are present in the explored intensity--energy plane (see text), so
  we restrict the axis ranges in the Figure for clarity. From
  outermost to innermost, the contours represent $\Delta\chi^2= -2.28,
  -5.99, -9.21, -13.816$,\, corresponding approximately to 68, 95, 99,
  and 99.9 per cent confidence regions.  }
\label{rescont}
\end{center}
\end{figure}

We start our analysis by considering joint fits to the XIS spectra in
the restricted 2--10~keV (FI) and 2--8~keV (BI) band. We first apply a
simple power law model modified by photoelectric absorption fixed at
$1.5 \times 10^{20}$~cm$^{-2}$ (the Galactic column in the line of
sight, Kaberla et al 2005). The data are well described by this simple
model ($\chi^2=665$ for 661 degrees of freedom, dof) and we obtain
$\Gamma = 1.70\pm 0.05$, and a 2--10~keV flux of $\simeq 5.1\times
10^{-12}$~erg~cm$^{-2}$~s$^{-1}$, corresponding to a 2--10~keV
unabsorbed luminosity of $\simeq 6.1\times 10^{44}$~erg~s$^{-1}$. No
striking residuals are present over the whole 2--10~keV band. However,
visual inspection of the residuals may not be the most efficient way to
look for spectral features. A more efficient visual representation of
the residuals is given in Fig.~\ref{rescont} where we use the method
outlined in Miniutti \& Fabian (2006) and run a Gaussian emission line
filter through the data varying its energy between 4~keV and 8~keV and
its intensity in the $\pm 1\times 10^{-4}$~ph~cm$^{-2}$~s$^{-1}$ range,
while keeping its width fixed at 1~eV, and we record the statistical
improvement. This procedure highlights that excess emission is present
around 6.4~keV in the rest--frame and the contours suggest a
statistical significance of more than 3$\sigma$. 

We then add a Gaussian emission line to our model with energy, width,
and normalisation free to vary. We obtain a significant improvement of
$\Delta\chi^2= -22$ for 3 dof, i.e. $\chi^2=643$ for 658 dof,
corresponding to a significance at the 99.9999 per cent level
according to a simple F--test (used to roughly estimate the true
statistical significance) and we measure a line energy of $6.40\pm
0.10$~keV, consistent with neutral Fe. The line is resolved, although
not extremely broad, with $\sigma = 0.17^{+0.10}_{-0.06}$~keV and has
an equivalent width EW$=60\pm 25 $~eV.

Since a narrow 6.4~keV Fe line is ubiquitous in the X--ray spectra of
AGN (see e.g. Bianchi et al. 2007), we then consider a model
comprising two Gaussian emission lines, one representing the putative
narrow line with energy $E\equiv 6.4$~keV and width $\sigma\equiv
1$~eV, the other (with all parameters free to vary) accounting for the
broader component. No improvement is however obtained with respect to
the best--fitting model described above (one broadish Gaussian only),
and the narrow line intensity is only an upper limit. Given that the
narrow line is not detected, we remove it from our spectral model. In
order to check whether the broad line width could be partially due to
the presence of a Compton shoulder (e.g. Matt 2002), we include an
Gaussian emission line with energy fixed at $6.3$~keV and $\sigma
\equiv 40$~eV. However, the Compton shoulder intensity tends to zero
and no change in the width of the 6.4~keV line is measured. We then
conclude that the Fe emission line we observe is broad with respect to
the XIS energy resolution and we remove the Compton shoulder component
from our model.

If the broad line energy,
width, and normalisation is allowed to vary between the three
detectors, no significant changes in the line parameters are obtained
for the two FI spectra, while the BI one provides consistent but less
constrained results (due to the lack of sensitivity above
7--8~keV). Thus the line is marginally resolved and line parameters are
consistent in the XIS detectors independently, with better constraints
coming from the two FI spectra.

The detected broad line has energy consistent with neutral Fe
K$\alpha$ emission. It is thus unlikely that its width could be
explained by Comptonization by partially ionized gas. Moreover, the
lower limit on the line width ($\sigma \geq 0.11$~keV) corresponds to
a FWHM of $\geq 1.2 \times 10^4$~km~s$^{-1}$, much larger than the
H$\beta$ FWHM~$=4.26\pm 1.25\times 10^3$~km~s$^{-1}$ (Grupe et
al. 2004), excluding that the Fe K$\alpha$ line comes from the broad
line regions and suggesting an origin in reflection off material much
closer to the central black hole. The natural production site is then
the accretion disc, where velocity and gravitational shifts can
contribute to broaden the line profile.  In this case the fluorescent
line emission must be associated with a full reflection spectrum which
also produces a soft excess for a wide range of ionization states
(e.g. Ross \& Fabian 1993; Zycki 1994; Nayakshin, Kazanas \& Kallman
2000; Rozanska et al. 2002; Ross \& Fabian 2005). We will consider
such model after extending the energy range of our spectral analysis.

\section{Including the soft X--ray data: the need for a multi--component spectral model}
\begin{figure}
\begin{center}
\includegraphics[width=0.35\textwidth,height=0.45\textwidth,angle=-90]{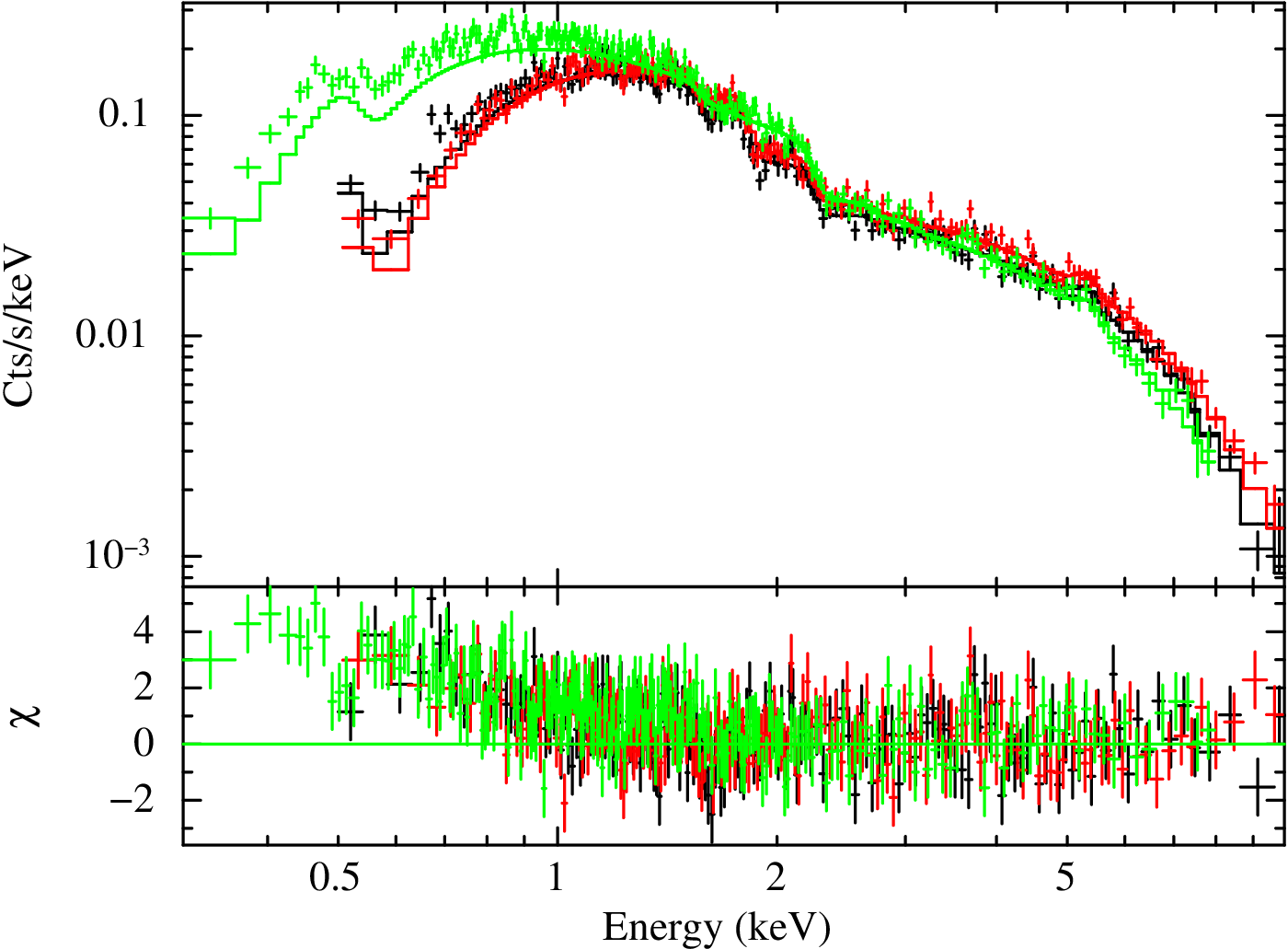}
                {\vspace{0.2cm}}
                \includegraphics[width=0.15\textwidth,height=0.45\textwidth,angle=-90]{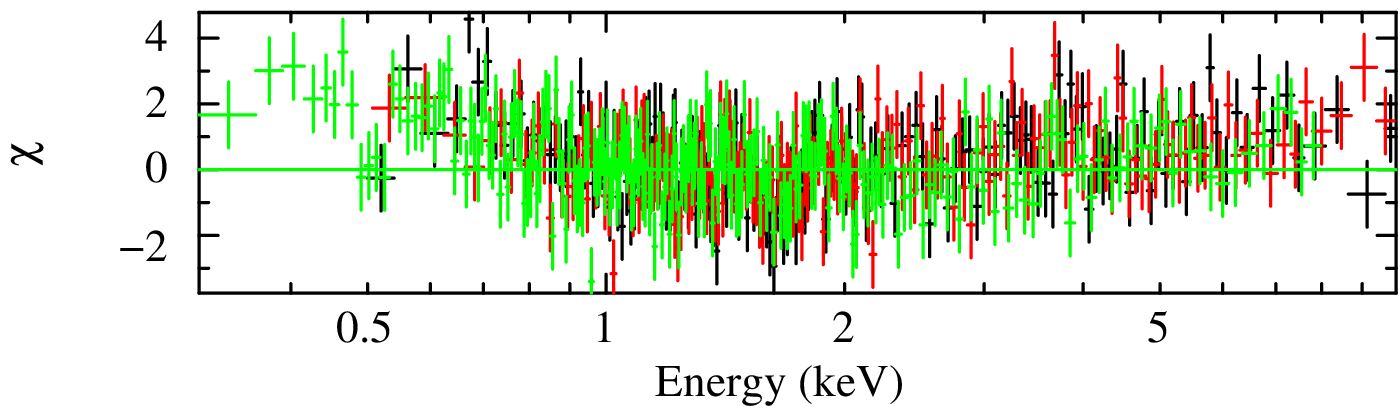}
                \caption{In the top panel we show the extrapolation to
                  soft X--rays of our simple 2--10~keV power law (plus
                  Gaussian) model which shows excess emission below
                  $\sim$1--2~keV. The BI spectrum is the one extending
                  to softer energies (green in the on--line version),
                  while the two FI spectra superimpose almost exactly
                  (XIS~0 in black and XIS~3 in red in the on--line
                  version).In the bottom panel, we show the residuals
                  left once the data have been re--fitted with the
                  same model. The model attempts to describe the soft
                  excess by steepening the power law photon index,
                  which however provides a worse description in the
                  hard band. We conclude that the soft excess is most
                  likely real and that the 0.3--10~keV spectral shape
                  cannot be described by a simple power law model. The
                  x--axis energy is in the observed frame.}
\label{sex}
\end{center}
\end{figure}

The extrapolation of our best--fitting model to the soft energies (down to
0.5~keV for the FI spectra and 0.3~keV for the BI), reveals a relatively weak
soft excess, as shown in the top panel of Fig.~\ref{sex}. 
Re--fitting the data with the same model provides a relatively poor
description of the data ($\chi^2=1672$ for 1466 dof) with a steeper
photon index than before ($\Gamma= 1.85\pm 0.04$). The residuals do not
look satisfactory and exhibit some curvature both in the hard and soft
band, suggesting that the soft excess is real and that the 0.3--10~keV
band spectrum has to be described with a multi--component model rather
than with a simple power law. This is shown in the bottom panel of
Fig.~\ref{sex}. We note that an absorption feature seems to be
present around 0.5~keV ($\sim$0.6~keV in the rest--frame). This
feature will be discussed when more appropriate spectral models will
be used.

\section{A self--consistent disc reflection model}

Given that we have evidence for a relatively broad Fe line we first
consider a two--component model comprising a power law continuum and
an X--ray reflection spectrum (from Ross \& Fabian 2005) which is
convolved with the relativistic kernel {\sc{kdblur}} to account for
the spectral distortions induced by the large orbital velocity and
strong gravity effects in the inner accretion disc. The relativistic
blurring parameters are the same as for the {\sc{laor xspec}} model
(Laor 1991), except that the spectrum to be affected is the full
X--ray reflection spectrum and not just a narrow emission line. We fix
the outer disc radius at its maximum allowed value of $400~r_g$ (where
$r_g=GM/c^2$), while the emissivity index $q$ (where the emissivity is
parametrised as $\epsilon (r)\propto r^{-q}$), the disc to observer
inclination $i$, and the inner disc radius $r_{\rm{in}}$ are free to
vary. As for the reflection model, the photon index of the irradiating
continuum and its high--energy cut--off are tied to the power law (the
cut--off is fixed at 100~keV, as assumed in the Ross \& Fabian
reflection grid of models), the Fe abundance is fixed at the Solar
value, while the ionization state of the reflector and the overall
reflection normalisation are free parameters.

We obtain a very good description of the 0.3--10~keV XIS data with
$\chi^2=1500$ for 1464 dof reproducing well the soft excess, Fe line,
and general spectral shape. In other words, the power law plus disc
reflection model provides a statistical improvement of $\Delta\chi^2=
-172$ for $\Delta dof = -2$ in the 0.3--10~keV band with respect to
the previous power law plus Gaussian model. Before discussing the
relevant parameters, we note that removing the relativistic blurring
kernel (i.e. fitting the data with a power law plus unblurred
reflection) produces a much worse description of the data with
$\Delta\chi^2= +40$ for $\Delta dof = +3$, i.e. the relativistic
blurring is required at the more than the 6$\sigma$ level. Since the
line model has now changed with respect to the broad Gaussian used
above, we re--consider the issue of whether a narrow 6.4~keV Fe
K$\alpha$ line is present by including a narrow Gaussian emission line
(with $E\equiv 6.4$~keV and $\sigma \equiv 1$~eV) in our model. The
narrow Fe line produces a further statistical improvement of
$\Delta\chi^2 = -10$ for 1 dof, i.e. the narrow line is now detected
at more than the 3$\sigma$ level with an equivalent width of $25\pm
13$~eV. The narrow line component was not required when using the
previous Gaussian model, which is understandable giving that
the broad line is now asymmetric, skewed towards low energies, and peaks at
slightly higher energy than 6.4~keV thus leaving some room for a
narrower line component (see e.g. the top panel of
Fig.~\ref{2comp}). A further absorption feature is seen in the
residuals (see also Fig.~\ref{sex}) and can be modelled with a narrow
($\sigma \equiv 1$~eV) Gaussian absorption line at $E_{\rm
  {abs}}\simeq 0.61$~keV. Despite its low equivalent width
($EW_{\rm{abs}}\simeq -8$~eV) and a somewhat uncertain identification
(no strong absorption feature is expected at 0.61~keV) we include it
in our model since it provides a statistical improvement
($\Delta\chi^2 = -10$ for 2 dof for a final result of $\chi^2 = 1480$
for 1461 dof) but we cannot exclude that it has an instrumental
origin.

\begin{table}
\begin{center}
  \caption{Best--fitting parameters for the power law plus
    relativistically blurred reflection model, also comprising an
    additional narrow Fe K$\alpha$ line with intensity $I_{\rm{NL}}$ and equivalent width
    $EW_{\rm{NL}}$ and a soft absorption line with energy
    $E_{\rm{abs}}$ and equivalent width $EW_{\rm{abs}}$. The
    superscript $^f$ denotes that the high--energy cut--off parameter
    is fixed at 100~keV.}
\begin{tabular}{lc}        
  \hline 
$\Gamma$ & $1.85 \pm 0.08 $ {\vspace{0.05cm}}\\ 
$E_{\rm c}$ [keV]& $100^f$ {\vspace{0.05cm}}\\ 
$i$ [degrees]& $33\pm 6${\vspace{0.05cm}}\\ 
$q$ & $4.1^{+5.3}_{-0.9}$ {\vspace{0.05cm}}\\ 
$r_{\rm{in}}$ [$r_g$] & $\leq 3.2$ {\vspace{0.05cm}}\\ 
$\xi_{\rm{ref}}$ [erg~cm~s$^{-1}$] & $40^{+70}_{-30}${\vspace{0.05cm}} \\ 
$I_{\rm{NL}}$ [ph~cm$^{-2}$~s$^{-1}$]& $(2.8\pm 1.4)\times 10^{-6}$ {\vspace{0.05cm}}\\
$EW_{\rm{NL}}$ [eV]& $25\pm 13$ {\vspace{0.05cm}}\\ 
$E_{\rm{abs}}$ [keV]& $0.61\pm 0.05$ {\vspace{0.05cm}}\\ 
$EW_{\rm{abs}}$ [eV]& $-8\pm 6$ {\vspace{0.05cm}}\\ 
$F_{2-10~keV}$ [erg~cm$^{-2}$~s$^{-1}$]& $\simeq 5.1\times 10^{-12}$ {\vspace{0.05cm}}\\ 
$L_{2-10~keV}$ [erg~s$^{-1}$]& $\simeq 6.0\times 10^{44}$ {\vspace{0.05cm}}\\ 
$\chi^2 /dof$ & 1480/1461\\ 
\hline
\end{tabular}
\end{center}
\label{tab1}
\end{table}

Our best--fitting model results in a 2--10~keV observed flux of
$\simeq 5.1\times 10^{-12}$~erg~cm$^{-2}$~s$^{-1}$, while the
2--10~keV unabsorbed luminosity is $\simeq 6.0\times
10^{44}$~erg~s$^{-1}$. In order to compute the source bolometric
luminosity from the 2--10~keV one, we adopt a bolometric correction of
50 (e.g. Marconi et al. 2004) which however refers to the intrinsic
2--10~keV luminosity (i.e. the power law luminosity $\sim
4.8\times 10^{44}$~eg~s$^{-1}$) leading to $L_{\rm Bol} \simeq 2.3
\times 10^{46}$~erg~s$^{-1}$. Assuming a black hole mass of $1.8\times
10^8~M_\odot$, RBS~1124 turns out to be particularly radiatively
efficient with an estimated Eddington ratio of $\sim 1$.

As for the most relevant best--fitting parameters, the ionization
state is constrained to be $\xi = 40^{+70}_{-30}$~erg~cm~s$^{-1}$ in
which range the dominant Fe line is the K$\alpha$ at 6.4~keV, in
agreement with our previous findings. We measure a photon index of
$\Gamma \simeq 1.85$, steeper than the power law model fitted in the
2--10~keV band because the model now also comprises the X--ray
reflection spectrum, which is very hard in the 2--10~keV band
(equivalent $\Gamma \simeq -0.1$). The relativistic blurring
best--fitting parameters are $i=33^\circ\pm 6^\circ$,
$q=4.1^{+5.3}_{-0.9}$, and $r_{\rm{in}}\leq 3.2~r_g$. We note that
the emissivity index is basically unconstrained and our result
translates effectively into $q\geq 3.2$ at the 90 per cent level.  A
summary of the inferred best--fitting parameters for the final model
is given in Table~1.

\begin{figure}
\begin{center}
\includegraphics[width=0.35\textwidth,height=0.45\textwidth,angle=-90]{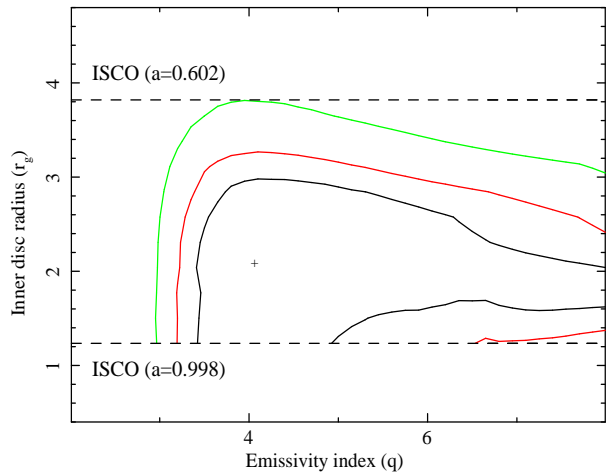}
\caption{$\Delta\chi^2= 2.3, 4.61, 9.21$ contours (corresponding to
  approximate 68.3, 90, and 99 per cent confidence regions) for the
  two most relevant relativistic parameters, namely the inner disc
  radius and the emissivity index. The two horizontal lines mark the
  ISCO and the associated black hole spin value. Under the assumption
  that the disc extends down to the ISCO, the black hole spin is
  constrained to be $a\geq 0.60$ at the 99 per cent level for two
  significant parameters.}
\label{rinco}
\end{center}
\end{figure}

Our result on the inner disc radius ( $r_{\rm{in}}\leq 3.2~r_g$) is
particularly intriguing. Under the standard assumption that the inner
disc radius coincides with the innermost stable circular orbit (ISCO)
of the black hole spacetime, $r_{\rm{in}}$ only depends on black hole
spin (with larger ISCO corresponding to smaller spin, Bardeen, Press
\& Teukolsky 1972). The upper limit we measure translates into a lower
limit of $a \geq 0.74$ for the black hole spin, suggesting that the
black hole in RBS~1124 is a rapidly rotating Kerr one. To investigate
in some more detail this important result, and to check upon the
possibility that it may be driven by a local rather than global
$\chi^2$ minimum, we have considered contour plots on the two most
relevant parameters, namely the inner disc radius and the emissivity
index. We consider these two parameters because they can in principle
be degenerate since both small inner radii and steep emissivity
profiles tend to produce broader and more redshifted line shapes,
while the disc inclination has a more independent effect, especially
for the intermediate $\simeq 33^\circ$ value we measure. The result of
this exercise is shown in Fig.~\ref{rinco}. As already suggested by
errors on single parameters, the emissivity index is basically only
constrained to be $q\geq 3$, but for any allowed value of $q$,
$r_{\rm{in}}$ never exceeds a rather small limiting radius. We
conclude that $r_{\rm{in}} \leq 3.8~r_g$ at the 99 per cent level,
which translates into a spin $a\geq 0.60$, still strongly pointing to
a rapidly rotating Kerr black hole solution (see however the
discussion in the following Section).

As a final remark, we point out that the reflection model we use does
not allow to compute the reflection fraction $R$ directly. However, it
can be approximately estimated using the relationship in Brenneman et
al. 2009 (see their Eq.~1) which, in our case, gives $R \sim 0.6$.
Note that the inferred reflection fraction is broadly consistent with
the relatively small measured broad plus narrow Fe line equivalent
width ($\simeq 85$~eV) for solar Fe abundance (George \& Fabian
1991). This amount of reflection should be partly associated with the
broad line ($R_{\rm{disc}}$) and partly with the narrow Fe line (most
likely originating in reflection off distant material such as the
obscuring torus, $ R_{\rm{torus}}$). The splitting between these two
components can be roughly estimated by comparing the lines EW, i.e
$EW_{\rm{broad}}/EW_{\rm{narrow}} \simeq R_{\rm{disc}}/R_{\rm{torus}}
~\simeq 2.4$ and thus $R_{\rm{disc}}\simeq 0.42$ and
$R_{\rm{torus}}\simeq 0.18$.

\subsection{On the reliability of the inferred $r_{\rm{in}}$ (and black hole spin)}

We must point out that our result on $r_{\rm{in}}$ is somewhat
surprising. According to the Fe line shape (see e.g. Fig.~2) and
Gaussian width ($\sigma \simeq 0.17$), the red wing of the line is not
extended to very low energies. It is difficult then to claim that the
Fe line shape implies $r_{\rm{in}} \leq 3.8$ because such as small
inner radius should result in a much more extended red wing (see
e.g. Fabian \& Miniutti 2009). The small reflection fraction we
measure implies that only the relatively narrow blue peak of the broad
Fe line is observable above the continuum, which is consistent with
the relatively small $\sigma$ we infer for the line, but it is clear
that the breadth of the Fe line red wing that is detectable above the
continuum is not extended enough to imply such a small inner disc
radius. A possible explanation for this apparent discrepancy is that
$r_{\rm{in}}$ is primarily driven by the need to describe a smooth
soft excess with the disc reflection model. Large $r_{\rm{in}}$ cannot
be able to account for a smooth soft excess because the
reflection soft X--ray lines (e.g. Oxygen, Fe~L, and others) would not
be sufficiently blurred.

To test whether the smaller inner disc radius is driven by the soft
excess smoothness we restrict the fitting interval to the 2--10~keV
band. We obtain a good representation of the data with $\chi^2=640$
for 655 dof. As for the most relevant parameter, we measure a much
larger inner disc radius of $r_{\rm{in}}\simeq 100~r_g$ (but basically
unconstrained). We thus conclude, as already suspected, that the Fe
line profile alone does not require the presence a disc with small
$r_{\rm in}$. It is only when the soft X--ray data are considered, and
under the assumption that the soft excess is due to disc reflection,
that a small inner disc radius is required by the data. The main
blurring parameters are then driven by the soft X--ray excess
smoothness and the Fe line only sets the overall reflection
normalisation and strength with respect to the continuum. Consistency
between the soft and Fe line data is achieved with a small inner disc
radius and a small reflection fraction, but the inferred $r_{\rm{in}}$
(and thus black hole spin $a$) is not driven by the Fe line
shape. Hence, the derived black hole spin lower limit ($a\geq 0.60$)
is in this case particularly interpretation--dependent and relies on
the idea that the soft excess is due to X--ray reflection from the
accretion disc. As mentioned before, the reflection fraction is lower
than standard, and this explains why the red wing of the broad line is
buried in the continuum, while only the relatively narrow blue peak of
the line is confidently detected above the continuum level.

The idea that the soft X--ray excess in AGN is due to disc reflection
is well motivated physically and is generally very successful when
tested on X--ray data (see e.g. Fabian et al. 2004; Ponti et al. 2006;
Miniutti et al. 2009b; Fabian et al. 2009), but it is nonetheless a
non unique interpretation. It should be however pointed out that the
detection of a short $\sim$30~s lag between the continuum and the soft
excess in the Narrow--Line Seyfert~1 galaxy 1H~0707--495 points
towards the idea that the soft excess is indeed part of a reprocessed
component, favouring the reflection interpretation in this extreme
source (Fabian et al. 2009). We also point out here that the high
Eddington ratio we infer from from the X--ray luminosity
suggests very strongly that a radiatively efficient inner disc/corona
structure is present, unless the black hole mass estimate of RBS~1124
(or the used bolometric correction) is wrong by at least one
order of magnitude, which seems unlikely. Hence the question is not,
in our opinion, whether the accretion flow extends down to/near the
ISCO, but rather whether we can constrain its inner radius and the
spacetime geometry precisely through X--ray data analysis in this
particular QSO.

\section{A Partial--covering alternative}

We have seen above that, under the assumption that the broadband
spectrum is shaped by a standard X--ray power law continuum plus its
reflection from the disc, a disc extending down to the ISCO of a
rapidly rotating Kerr black hole is required by the
data. However, as mentioned above, our result depends strongly on the
assumption that the soft X--ray excess is due to disc reflection.

We consider here, as a possible alternative to the reflection model
presented above, a scenario in which the 0.3--10~keV spectrum of
RBS~1124 is described by a simple power law seen however through a
column of partially ionized gas which covers only partially the X--ray
emitting region. Our model comprises Galactic absorption, a power law,
a Gaussian emission line (with energy, width, and normalisation free
to vary), and a partially ionized partial covering model (the
{\sc{zxipcf}} model in {\sc{xspec}}) characterised by three free
parameters (column density, ionization state, and covering
fraction). We apply the model to the 0.3--10~keV XIS data and we
obtain a very good description of the spectral shape, statistically
equivalent to the one discussed above ($\chi^2=1493$ for 1463 dof). As in the
previous model, adding a soft absorption line at $\sim$0.61~keV
improves the fit to $\chi^2=1482$ for 1461 dof, a result which cannot
be statistically distinguished from the reflection--based one
($\chi^2=1480$ for the same number of dof). A summary of the
best--fitting model parameters is given in Table~2.

The partial coverer has a column density N$_{\rm
  H}=6.0^{+3.0}_{-2.0}\times 10^{22}$~cm$^{-2}$, ionization parameter
$\log\xi = -0.5^{+0.7}_{-2.5}$, and covers $36 \pm 7$ per cent of the
X--ray emitting region, while the intrinsic spectral shape is
characterised by $\Gamma = 2.01\pm 0.07$. As for the Gaussian emission
line, we measure $E=6.4\pm 0.1$~keV, $\sigma =
0.16^{+0.14}_{-0.06}$~keV for a line EW of $58\pm 30$~eV. In the
partial covering scenario, the unabsorbed 2--10~keV luminosity is
$\sim 7.3\times 10^{44}$~erg~s$^{-1}$ which implies an Eddington ratio
of $\sim 1.5$, significantly higher than that estimated with the
previous reflection model ($\sim 1$). Again, such high Eddington
ratio tells us that radiatively efficient accretion is taking place even when a partial covering scenario is considered, so
that a disc/corona structure close to the ISCO is most likely 
present. The question is whether the inner accretion flow is visible
(e.g. through X--ray reflection) or masked by absorption.

\begin{table}
\begin{center}
  \caption{Best--fitting parameters for the partial covering model
    (with covering fraction $Cf$), also comprising a resolved Fe
    emission line and the soft absorption line already present in the
    reflection model (see Table~1).}
\begin{tabular}{lc}        
  \hline $\Gamma$ & $2.01 \pm 0.07$ {\vspace{0.05cm}}\\ $N_{\rm H}$
         [cm$^{-2}$] & $ 6.0^{+3.0}_{-2.0} \times 10^{22}$
         {\vspace{0.05cm}}\\ $\log \xi$ &
         $-0.5^{+0.7}_{-2.5}${\vspace{0.05cm}} \\ $Cf$ & $0.36\pm
         0.07${\vspace{0.05cm}} \\ $E_{\rm{Fe}}$ [keV]& $6.40\pm 0.10$
         {\vspace{0.05cm}}\\ $\sigma_{\rm{Fe}}$ [keV]&
         $0.16^{+0.14}_{-0.06}$ {\vspace{0.05cm}}\\ 
$I_{\rm{Fe}}$ [ph~cm$^{-2}$~s$^{-1}$]& $(5.7\pm 2.7)\times 10^{-6}$ {\vspace{0.05cm}}\\ 
$EW_{\rm{Fe}}$ [eV]& $58\pm 30$ {\vspace{0.05cm}}\\ 
$E_{\rm{abs}}$ [keV]&
         $0.61\pm 0.06$ {\vspace{0.05cm}}\\ $EW_{\rm{abs}}$
         [eV]& $-8\pm 6$ {\vspace{0.05cm}}\\ $L_{2-10~keV}$
         [erg~s$^{-1}$]& $\simeq 7.3\times 10^{44}$
         {\vspace{0.05cm}}\\ $\chi^2 /dof$ & 1482/1461\\ \hline
\end{tabular}
\end{center}
\label{tab1}
\end{table}

The absorber ionization parameter is low and no strong absorption
lines are predicted by the model so that no velocity shifts can be
measured. The low ionization state is consistent with the idea that
the neutral Fe K$\alpha$ line we detect originates from reflection
and/or transmission in the gas responsible for the partial
covering. If the observed Fe line can be produced by the absorber, the
model would have the same satisfactory level of self--consistency than
the reflection model presented above with only two components
(continuum and either disc reflection or partial covering absorption)
accounting for the whole 0.3--10~keV spectral shape. We first check
the significance of the resolved width of the Fe line by forcing it to
be unresolved ($\sigma \equiv 1$~eV) and by refitting the data. This
procedure provides a $\Delta\chi^2=+6.4$ with respect to the case of a
broad line, i.e. the Fe line is resolved at the 98.8 per cent level
according to the F--test. 

Assuming that the Fe line is associated with the absorber, its FWHM
can be used to constrain the absorber location $R_{\rm{abs}}$ under
the hypothesis that the absorbing gas moves with isotropic Keplerian
velocity $v_{\rm{abs}}=\sqrt{3}/2$~FWHM. On the other hand, the
ionization state of the absorber is defined in the {\sc{zxipcf}} model
as $\xi = L\,\Delta R_{\rm{abs}}\, N_{\rm H}^{-1}\, R_{\rm{abs}}^{-2}$
(where $L$ is the central engine luminosity\footnote{In the partial
  covering model we use the intrinsic luminosity L, defined as the
  continuum luminosity in the 1--1000~Ry range. We estimate L by
  extrapolating our unabsorbed power law continuum model in that
  range.} and $\Delta R_{\rm{abs}}$ is the absorber thickness) and can
thus be used to constrain the absorber relative thickness $\Delta R/R
= \xi\, R_{\rm{abs}}\, N_{\rm H} / L$. By using the available upper
values of $\xi, R_{\rm{abs}}, N_{\rm H}$ and lower value of the
intrinsic luminosity, we estimate $\Delta R/R \leq 6 \times 10^{-6}$
for a black hole mass of $1.8\times 10^8~M_\odot$. The inferred
$\Delta R/R$ is implausibly small, indicating that our basic
assumption, namely that the Fe line is produced by the absorber, is
not correct. We are thus left with an unknown origin for the Fe line
we observe, so that the partial covering scenario is less
self--consistent than the reflection one and does require an
additional emission component to account for the resolved Fe K$\alpha$
emission line. Moreover, since the Fe line is resolved, the additional
component cannot be the standard obscuring torus of AGN unification
schemes because such material is located at the $\sim$~pc scale, thus
producing unresolved emission lines.

\section{The broadband 0.3--30~keV spectrum}

\begin{figure}
\begin{center}
\includegraphics[width=0.35\textwidth,height=0.45\textwidth,angle=-90]{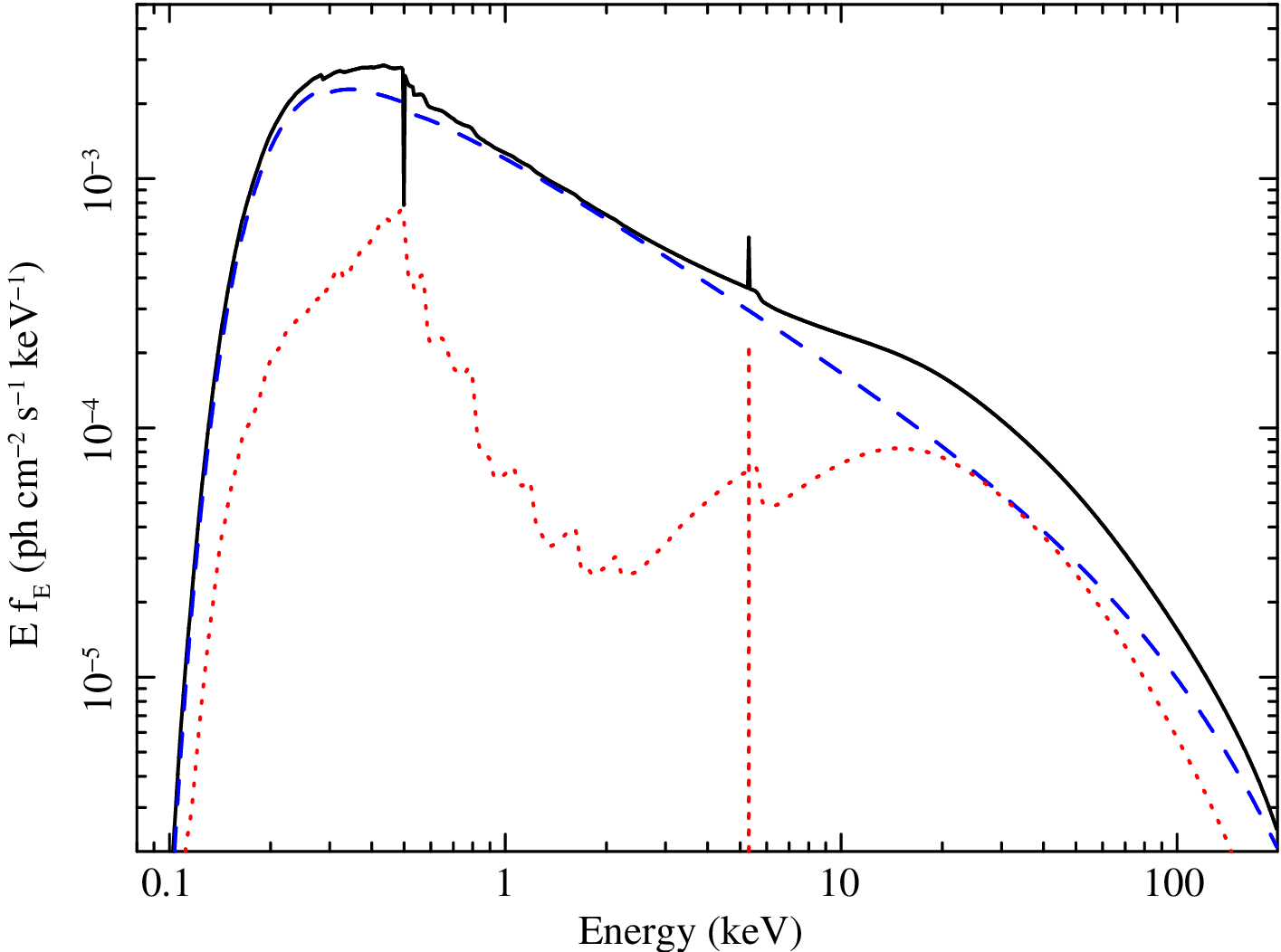}
{\vspace{0.5cm}}
\includegraphics[width=0.35\textwidth,height=0.45\textwidth,angle=-90]{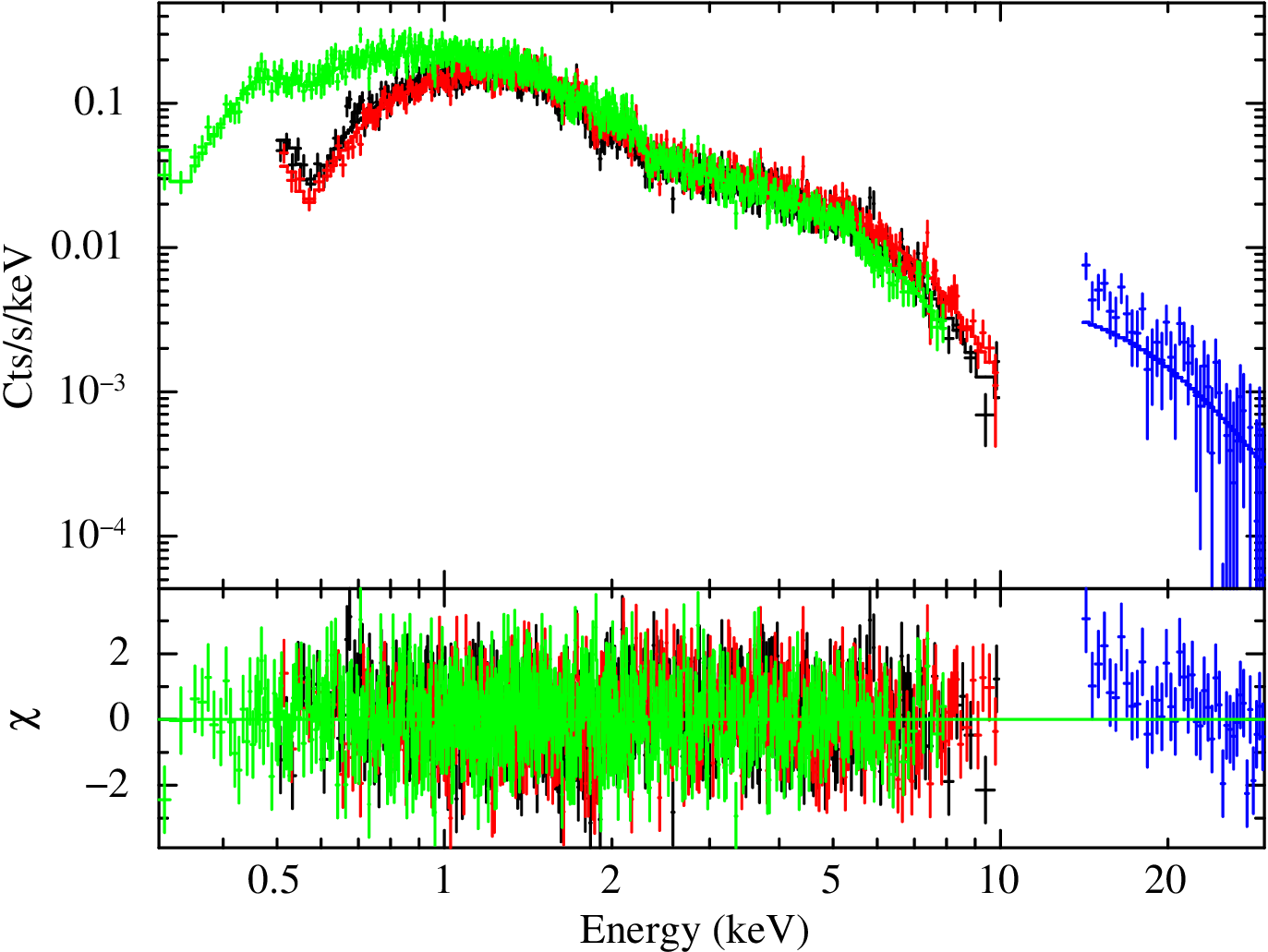}
{\vspace{0.5cm}}
\includegraphics[width=0.15\textwidth,height=0.45\textwidth,angle=-90]{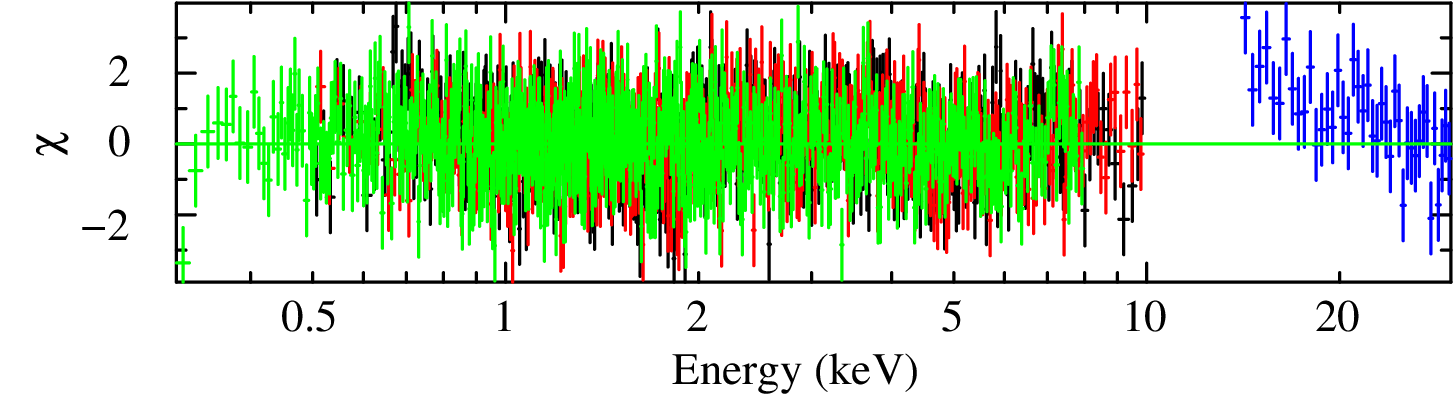}
\caption{In the top panel we show the best--fitting two--component
  model (power law plus disc reflection) including also the unresolved
  narrow Fe K$\alpha$ emission line at 6.4~keV and the $\sim$0.61~keV
  absorption line. In the middle panel, we show XIS and PIN data,
  best--fitting reflection model, and residuals. In the bottom panel,
  we show the residuals for the best--fitting partial covering
  model. The same detector color scheme as in Fig.~\ref{sex} is used,
  and the PIN data are shown in blue in the on--line version. Note the
  weak excess in the PIN 14--30~keV band for both adopted models (see
  text for discussion). The x--axis is in the observed frame.}
\label{2comp}
\end{center}
\end{figure}

The two competing models presented above (disc reflection and partial
covering) make somewhat different predictions above 10~keV. We then
consider the HXD/PIN data in the 14--30~keV range where the source is
detected with good statistics. The PIN signal in that band is about
8.8 per cent of the combined non X--ray and Cosmic X--ray backgrounds
(NXB and CXB). We introduce a cross--normalisation PIN/XIS~0 constant
of $1.18\pm 0.02$ as discussed in Maeda et al
(2008\footnote{JX--ISAS--SUZAKU--MEMO--2008--06.}). Plain
extrapolation of the 0.3--10~keV best--fitting models gives
$\chi^2=1539$ and $\chi^2=1603$ for the reflection and partial
covering models respectively for the same number of dof
(1502). Refitting the data does not improve much the reflection model
statistics ($\chi^2=1535$) and we do not observe any significant
changes in the parameters, while a more substantial improvement is
obtained in the partial covering case ($\chi^2=1562$) resulting in a
covering fraction of $\sim$55 per cent with much larger N$_{\rm
  H}=(1.2\pm 0.2) \times 10^{24}$~cm$^{-2}$ and $\log\xi = 2.3\pm
0.3$. Note however that in this case the intrinsic 2--10~keV
luminosity becomes $\simeq 1.1\times 10^{45}$~erg~s$^{-1}$,
corresponding to a perhaps implausibly high super--Eddington ratio of
$\simeq 2.4$. In Fig.~\ref{2comp} we show the best--fitting reflection
model (top panel), the data, model, and residuals for the reflection
model (middle panel), and the residuals for the best--fitting partial
covering one (bottom panel). 

It should be stressed that, even for the best--fitting reflection
model, the 14--30~keV PIN data are underestimated (see bottom panel of
Fig.~\ref{2comp}). By assuming a $\Gamma\equiv 2$ power law fitted
only to the PIN data, we measure an excess flux of $\simeq 2\times
10^{-12}$~erg~cm$^{-2}$~s$^{-1}$ at the 99.1 per cent confidence level
according to the F--test. We have inspected the quasi--simultaneous
{\it Swift} XRT/PC images (see Section 8 below) to look for possible
contaminating hard X--ray sources, but we only detected two sources
above 3~keV in a 9$\times$9~arcmin region around RBS~1124 and they
have a cumulative extrapolated 14--30~keV flux (assuming $\Gamma\equiv
2$) of $\simeq 2.5\times 10^{-13}$~erg~cm$^{-2}$~s$^{-1}$, i.e. about
one order of magnitude below the observed excess. The PIN field of
view is larger (about 34$\times$34~arcmin), but an archive--based
search did not provide any X--ray source that could explain the
observed PIN excess.

We cannot exclude that a variable hard X--ray source is contaminating
the PIN data and therefore we do not attach to much significance to
the better statistical result obtained with the reflection model when
fitting the PIN data. However, as mentioned, the best--fitting partial
covering model in the 0.3--30~keV band would require that RBS~1124 is
radiating at a super--Eddington rate, which seems unlikely.

\begin{figure}
\begin{center}
\includegraphics[width=0.35\textwidth,height=0.45\textwidth,angle=-90]{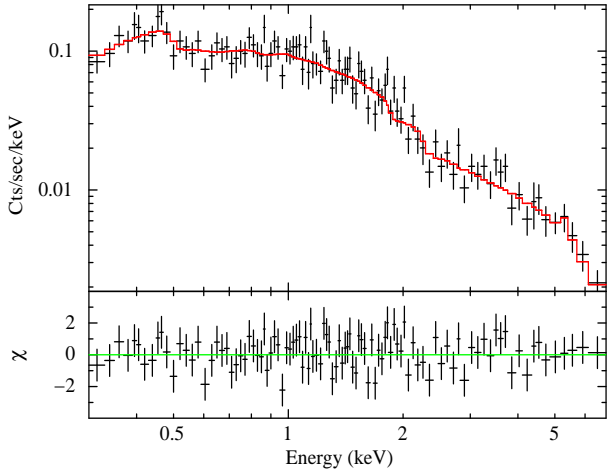}
\caption{The co--added {\it Swift} XRT/PC spectrum (14~ks of total
  exposure) together with the best--fitting reflection model and
  residuals. An identical statistical quality is reached when the
  partial covering model is used instead.  }
\label{xrt}
\end{center}
\end{figure}

\section{Long--term spectral variability}

As mentioned, RBS~1124 was detected in the RASS on 1990--10--26 and by
the {\it XMM--Newton} Slew Survey on 2006--11--25. In 2007, it was
also observed 4 times by {\it Swift} from May 24 to July 21 and data
from the XRT/PC are available.

The RASS spectrum is well represented by a simple $\Gamma \simeq 2.38$
power law, affected by Galactic absorption resulting in a 0.2--2~keV
flux of $7.6\times 10^{-12}$~erg~cm$^{-2}$~s$^{-1}$ (Grupe et
al. 2001). By assuming the same spectral shape, the {\it XMM--Newton}
Slew Survey flux is $6.5\times 10^{-12}$~erg~cm$^{-2}$~s$^{-1}$ in the
same band, implying limited flux variability in the two observations
separated by $\sim$16 years. For comparison, a similar fit to the {\it
  Suzaku} data (taken on 2007--04--14) yields a slightly harder slope
$\Gamma=1.95\pm 0.03$ and a much smaller 0.2--2~keV flux of $(4.07 \pm
0.07) \times 10^{-12}$~erg~cm$^{-2}$~s$^{-1}$. We conclude that
RBS~1124 was at its brightest during the RASS and has now dropped in
flux by slightly less than 50 per cent with some evidence for
associate hardening of the soft X--ray spectrum (from
$\Gamma_{\rm{RASS}} = 2.38 \pm 0.08$ to $\Gamma_{\rm{Suzaku}} =
1.95\pm 0.03$).

The {\it Swift} data are much closer in time to the {\it Suzaku}
ones (a few months separation) than the RASS. We first consider the
four XRT/PC spectra separately and apply a simple absorbed power law
model. This results in consistent spectral slope ($\Gamma \simeq 1.9$)
and 0.3--10~keV flux ($\simeq 8.2 \times
10^{-12}$~erg~cm$^{-2}$~s$^{-1}$) in all pointings. We thus produce a
co--added spectrum of all the XRT/PC observations resulting in a total
exposure of $\sim$14~ks. Fitting the data in the 0.3--2~keV band
provides $\Gamma = 1.9\pm 0.1$ and an extrapolated 0.2--2~keV flux of
$(4.3\pm 0.2)\times 10^{-12}$~erg~cm$^{-2}$~s$^{-1}$, broadly
consistent with the {\it Suzaku} results in the same band.

The XRT/PC data do not have enough quality to constrain the best--fitting
models any better than {\it Suzaku}, but we nevertheless apply our
best--fitting {\it Suzaku} models to the XRT/PC data to check for
consistency. Both the reflection and partial covering models provide a
good statistical description of the data with $\chi^2 =100$ for 93
dof, and no significant change in the best--fitting parameters is
observed in either model with respect to the results in Table~1 and
2. In Fig.~\ref{xrt} we show the co--added XRT/PC spectrum together with
the best--fitting reflection model and residuals.

\section{Discussion}

The 0.3--10~keV X--ray spectrum of RBS~1124 can be described by i) a
two--component model comprising a power law continuum and its
reflection off the accretion disc, or by ii) a partial covering model
in which an additional emission component is required to account for
the detected resolved Fe line (we note however that the line is
resolved at the 98.8 per cent confidence level). The two competing
models are statistically equivalent, but the reflection model appears to be more
self--consistent because it can explain all spectral features (namely
the general spectral shape, the soft excess, and the Fe line)
simultaneously. Moreover, when the models are applied to the
14--30~keV PIN data as well, the reflection model provides a much
better statistical description of the data and, most importantly, the
best--fitting partial covering one would imply a perhaps implausibly
high Eddington ratio of $\simeq 2.4$ (but see Section 7 for the
uncertainties associated with the PIN data).

On the other hand, the presence of material potentially able to
partially cover the X--ray source in the inner regions of accreting
supermassive black holes is well established in a few cases. The most
convincing cases are perhaps those in which absorber variability has
been observed on relatively short timescales (e.g. Lamer, Uttley \&
McHardy 2003; Risaliti et al. 2009; Bianchi et al. 2009; see also
Turner \& Miller 2009 and references therein for a review on this
active research topic). In most of the cases, the absorber variability
properties point to an identification of the absorber with broad line
region clouds or with disc winds originating from a similar
location. As discussed above, we cannot exclude that partial covering
from (for instance) broad line region clouds is occurring in RBS~1124,
but the observed Fe line must come from a distinct region, closer to
the central black hole, so that we need additional components in the
partial covering scenario, while the reflection model is more
self--consistent.

In the following we discuss the reflection scenario as the most
plausible description of the X--ray properties of RBS~1124 because of
its greater degree of self--consistency and because it does not invoke
any additional physical component with respect to the simplest
possible scenario only comprising an accretion disc and an X--ray
emitting corona above it. 


The most
important result of our reflection--based analysis is that the inner
disc radius of the reflector (disc) is tightly constrained to be
$r_{\rm{in}} \leq 3.8~r_g$ at the 99 per cent level. As discussed
above, this result is not driven by the relativistic Fe line shape,
but rather by the soft excess smoothness which requires a high degree
of relativistic blurring to mask the otherwise sharp soft X--ray
emission lines reflected off the disc. Our results must then be
considered as model--dependent or, better, as
interpretation--dependent because it relies on the idea that soft
excesses are associated with the soft part of a partially ionized
reflection spectrum from the accretion disc. If our interpretation is
correct, the upper limit on $r_{\rm in}$ translates into a lower limit
for the black hole spin of $a\geq 0.60$, strongly suggesting that the
black hole powering RBS~1124 is spinning rapidly. A small inner disc
radius is also consistent with the high Eddington ratio ($\sim 1$)
we obtain by applying standard bolometric corrections to the 2--10~keV
intrinsic luminosity. It would be very surprising if a black hole
radiating away its Eddington luminosity would not
possess a disc/corona structure extending down close to the black hole
event horizon. The relatively high lower limit on the black hole spin
may indicate that black hole growth in this object does not occur
through chaotic accretion episodes (and mergers) and favours instead
coherent accretion scenarios or recent mergers, as discussed in detail
e.g. in Berti \& Volonteri (2008).

RBS~1124 is a moderate redshift radio--quiet optically luminous quasar
whose X--ray properties can be compared with the well known class of
PG quasars. When a simple power law model is used, RBS~1124 hard
2--10~keV X--ray slope of $\Gamma_{\rm h}=1.70\pm 0.05$ is broadly
consistent with the typical hard slope of PG quasars ($<\Gamma_{\rm
  h}> \simeq 1.87$ with a dispersion $\sigma \simeq 0.36$; Piconcelli
et al. 2005). On the other hand, if a simple power law model is
applied in the soft band, RBS~1124 soft 0.5--2~keV X--ray spectral
slope is $\Gamma_{\rm s}=1.95\pm 0.03$, significantly harder than the
average PG quasars value ($<\Gamma_{\rm s}> \simeq 2.73$ with a
dispersion $\sigma \simeq 0.28$). This indicates a relatively small
soft excess in RBS~1124 with respect to the average PG quasar. In
fact, if a phenomenological blackbody (BB) plus power law (PL) model
is applied in the 0.3--10~keV Suzaku band, the ratio of the BB
luminosity to the total one in the 0.5--2~keV band is of $\sim$15 per
cent, about half of that typically observed (see e.g. Miniutti et
al. 2009a). 

In the framework of our two--component model, the soft excess strength
is determined by the disc reflection ionization state and, most
importantly, by the relative intensity of the reflection component
with respect to the direct power law continuum (i.e. the reflection
fraction). In the case of RBS~1124, the inferred disc reflection
fraction is small ($R_{\rm{disc}}\sim 0.42$) and explains why the soft
excess in this source is smaller than typical (in the standard case
the disc is assumed to cover half of the sky as seen by the primary
X--ray source, corresponding to $R_{\rm{disc}}\sim 1$). The reflector
could subtend a smaller area if the disc was truncated at large radii,
but this picture is not consistent with i) the measured value of
$r_{\rm in}$ and, perhaps more importantly, ii) the high Eddington
ratio of RBS~1124 and its radiatively efficient nature. In fact,
truncated disc are only invoked in low--luminosity and/or radio--loud
AGN and in Galactic black holes in the hard state, i.e. when the flow
radiative output is highly sub--Eddington ($L/L_{\rm{Edd}} \leq 0.02$
or so).

Alternatively, the X--ray corona, which provides the hard X--ray
photons irradiating the reflector, might be outflowing with mildly
relativistic velocity, as suggested by Beloborodov (1999) to explain
the low reflection fraction observed in Galactic black holes in the
hard state. An outflow velocity of the order of 0.25~c would reduce
$R$ from $\sim 1$ to $\sim 0.4$ due to aberration. Since RBS~1124 is
radio-quiet, the corona is  unlikely to
be the base of a mature relativistic jet, but could possibly be
associated with a failed jet. Note also that, under the assumption
that the X--ray continuum is due to Comptonization and for
$\beta=v/c=0.25$ corresponding to a Lorentz factor
$\gamma=(1-\beta^2)^{-1/2}=1.033$, the Beloborodov model predicts a
photon index $\Gamma \simeq 2\,[\gamma\,(1+\beta)]^{-0.3}\simeq 1.85$
in excellent agreement with the best--fitting value in the reflection
model case ($\Gamma=1.85\pm 0.08$ see Table~1). Moreover, assuming
vertical motion with $\beta=0.25$, mild beaming of the X--ray emission
would also reduce the intrinsic 2--10~keV luminosity by a factor
$D^{3}=[\gamma\,(1-\beta\,cos i)]^{-3}\simeq 0.53$ (where $i$ the
observer--velocity inclination) leading to $L_{\rm{Bol}} \simeq
1.2\times 10^{46}$~erg~s$^{-1}$ and to a reduced Eddington ratio of
$\simeq 0.5$. 
Note that partial covering does not suggest beaming of
the X--ray emission so that the estimated bolometric luminosity in
this case is much less consistent with the literature data, predicting
an Eddington ratio of 1.5 or 2.4 depending on the adopted
best--fitting model in the 0.3--10~keV or 0.3--30~keV band. This
further supports our interpretation of the X--ray properties of
RBS~1124 in terms of reflection from the inner accretion disc of a
mildly beamed X--ray continuum.

\section*{Acknowledgements}

This research has made use of data obtained from the {\it Suzaku}
satellite, a collaborative mission between the space agencies of Japan
(JAXA) and the USA (NASA).  GM thanks the Ministerio de Ciencia e
Innovaci\'on and CSIC for support through a Ram\'on y Cajal
contract. CV and SB thank the Italian Space Agency (ASI) for support through
grants ASI--INAF I/023/05/0 and ASI I/088/06/0. We also thank Roberto
Gilli for useful discussion on the uncertainty of the CXB flux in the
hard X--ray band.

\end{document}